\documentclass[conference]{IEEEtran}
\usepackage{cite}
\usepackage{amsmath,amssymb,amsfonts}
\usepackage{algorithmic}
\usepackage{graphicx}
\usepackage{textcomp}
\usepackage{xcolor}

\def\BibTeX{{\rm B\kern-.05em{\sc i\kern-.025em b}\kern-.08em
    T\kern-.1667em\lower.7ex\hbox{E}\kern-.125emX}}
\begin{document}

\title{On the dynamics of affective states during play and the role of confusion\\
{\footnotesize Short Paper Track}}

\author{
\IEEEauthorblockN{1\textsuperscript{st} Thomas Vase Schultz Volden}
\IEEEauthorblockA{\textit{Center for Digital Play - BrAIn Lab} \\
\textit{IT University of Copenhagen}\\
Copenhagen, Denmark \\
thvo@itu.dk \\
0009-0009-2177-5482}
\and
\IEEEauthorblockN{2\textsuperscript{nd} Oleg Jarma Montoya}
\IEEEauthorblockA{\textit{IT University of Copenhagen}\\
Copenhagen, Denmark \\
oleja@itu.dk \\
0009-0008-1036-5069}
\and
\IEEEauthorblockN{3\textsuperscript{rd} Paolo Burelli}
\IEEEauthorblockA{\textit{Center for Digital Play - BrAIn Lab} \\
\textit{IT University of Copenhagen}\\
Copenhagen, Denmark \\
pabu@itu.dk \\
0000-0003-2804-9028}
\and
\IEEEauthorblockN{4\textsuperscript{th} Marco Scirea}
\IEEEauthorblockA{\textit{SDU Metaverse Lab} \\
\textit{University of Southern Denmark}\\
Odense, Denmark \\
msc@mmmi.sdu.dk \\
0000-0003-4908-0526}
}

\maketitle

\begin{abstract}
Video game designers often view confusion as undesirable, yet it is inevitable, as new players must adapt to new interfaces and mechanics in an increasingly varied and innovative game market, which is more popular than ever.
Research suggests that confusion can contribute to a positive experience, potentially motivating players to learn.
The state of confusion in video games should be further investigated to gain more insight into the learning experience of play and how it affects the player experience.
In this article, we design a study to collect learning-related affects for users playing a game prototype that intentionally confuses the player.
We assess the gathered affects against a complex learning model, affirming that, in specific instances, the player experience aligns with the learning experiences. 
Moreover, we identify correlations between these affects and the Player Experience Inventory constructs, particularly concerning flow experiences.
\end{abstract}


\section{Introduction}
Game designers tend to optimise engagement by reducing confusion through clear, achievable goals and feedback~\cite{sweetser_gameflow_2005, chen_flow_2007}.

\begin{figure}
    \centering
    \includegraphics[width=0.5\linewidth]{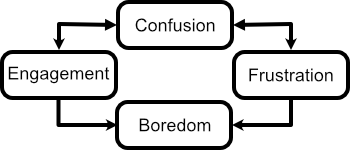}
    \caption{State diagram showing the Model of Affect Dynamics. States represent affective states, and the edges represent the likely transition.}
    \label{fig:MAD}
\end{figure}

However, studies suggest that confusion can be a positive experience that elicits excitement and can even improve engagement~\cite{arguel_puzzle-solving_2019}.
Studies of the Model of Affect Dynamics have shown predictable transitional dynamics between affects, shown in fig. \ref{fig:MAD}, for students who engage in complex learning with intelligent tutor software~\cite{dmello_dynamics_2012}.
This has been supported in puzzle games~\cite{arguel_puzzle-solving_2019} and educational games for young children~\cite{volden_childrens_2024}.

In the study presented in this paper, we collect player experience by frequent self-reports of one or two affects during play and test if these affects adhere to the Model of Affect Dynamics (fig. \ref{fig:MAD}).
Finally, we compare the experience with the well-established Player Experience Inventory survey tool, called MiniPXI~\cite{haider_minipxi_2022}, to correlate self-reported affects during play with constructs related to the experience of the flow state~\cite{csikszentmihalyi_toward_2014}.

\subsection{Literature Review}

Video game designers use components such as mechanics, dynamics, and aesthetics (MDA) to create a dynamic system that affords control and responds to players in a way that produces an intended player experience, such as a social fellowship or drama~\cite{hunicke_mda_2004, salen_rules_2004}.
Flow experience is considered the optimal player experience~\cite{chen_flow_2007}, referring to the experience of the flow state~\cite{csikszentmihalyi_toward_2014}, and can be obtained through clear goals and feedback~\cite{sweetser_gameflow_2005}.

Game designers use survey tools such as the Player Experience Inventory (PXI)~\cite{abeele_development_2020} to collect constructs of player experience (e.g. audiovisual appeal, challenge, ease of control, clarity of goals, progress feedback, autonomy, curiosity, immersion, mastery, meaning, and enjoyment) related to the MDA design components.
PXI has been used to rank the player experience for more than 360 games\footnote{PXI Bench: https://playerexperienceinventory.org/bdata}, and there is a convenient short version available called MiniPXI \cite{haider_minipxi_2022}.
%
The flow experience can be classified in the PXI by the constructs of \emph{immersion} and \emph{mastery}~\cite{haider_minipxi_2022}.
%
The experience of learning complex subjects such as maths has been studied in intelligent tutor systems, using self-reports \cite{dmello_dynamics_2012}, observed expressions, and data mining techniques~\cite{baker_towards_2012}.
They report four affects that students tend to experience: engagement/concentration, confusion, frustration, and boredom, and the transitions between these affects have been shown to follow a predictable pattern when students encounter new/contradictory information or impasses.
In these cases, students tend to transition from engagement to confusion as they process information or try to overcome an impasse~\cite{dmello_dynamics_2012}.
Confusion is described as a state of cognitive disequilibrium, which means that a person's internal model of the world is challenged and that the person is intrinsically motivated to regain equilibrium~\cite[p.31]{jeanne_ellis_ormrod_educational_2019}.
Confusion can be seen as a signal of learning because it signifies that a person's internal model of the world was challenged~\cite{dmello_confusion_2014}.
Confusion has been shown to be a positive or negative experience, based on how the student manages to adapt to the information and overcome the impasse~\cite{dmello_confusion_2014, arguel_puzzle-solving_2019}. A struggling student tends to go from productive confusion to hopeless confusion, leading to frustration, thereby making confusion a predecessor of frustration.
Students who managed to adapt tend to return to engagement with a positive experience\cite{arguel_puzzle-solving_2019} and increased emotional arousal, which benefits cognitive engagement~\cite{arguel_inside_2017}.
Boredom in these studies is related to the act of disengaging from the system, which usually occurs if a student is stuck in a frustrating, hopeless situation~\cite{dmello_dynamics_2012}, but studies have also shown that students can go from engagement to boredom and that confusion can be a preventive measure of boredom~\cite{arguel_inside_2017}.%
The main difference between Flow theory and the Model of Affect Dynamics is whether the subject experiences confusion before frustration.
Puzzle elements and learning games, which contain cognitive challenges, have been shown to produce an experience that seems to adhere to the Model of Affect Dynamics~\cite{arguel_puzzle-solving_2019, volden_childrens_2024}; however, the model remains to be applied to games that are not abstract puzzles or learning games.

\section{The study}
We designed a study in which players were asked to play-test a game prototype featuring a novel mix of mechanics\footnote{game prototype: https://tvoldenitu.itch.io/coltag}, to test if the experience adheres to the Model of Affect Dynamics. 
During the play test, the game paused every minute while the player selected one or two affects (engagement, confusion, frustration, boredom, and neutral) that best describe their past minute experience.
Given the limited transitions in the Model of Affect Dynamics (fig. \ref{fig:MAD}), players who adhere to the model should be unlikely to report two non-adjacent affects together; therefore, we formulated the following two hypotheses:
\begin{itemize} 
    \item[H1] Players are unlikely to report frustration and engagement at the same time.
    \item[H2] Players are unlikely to report confusion and boredom at the same time.
\end{itemize}

The game was designed as a mix of game mechanics from adventure games~\cite{karhulahti_mechanicaesthetic_2011} and a game called Story Teller~\cite{benmergui_storyteller_2023}.
The aesthetics were inspired by a classic adventure game series called Riddle School~\cite{jonbro_riddle_2006-1}, where puzzles are presented in the context of a game world.
Like many adventure games, the player has an inventory of tools which the player can use to overcome obstacles.
However, the player cannot apply the tool directly to items in the game world to get instant feedback, but will have to plan their action and execute the plan before receiving feedback.
To plan actions, the player has to create scenes using a tool called \emph{scene tool} from their inventory.
At the top of the screen, there is a planning bar consisting of four rectangular slots. Players can organise actions by dragging scene items into these empty slots and then interact with the environment by moving their player avatar or items into the scene.
Once the player has assigned scenes and actions to the planner, they can execute the plan by clicking the red play button on the planner bar.
The scenes are performed sequentially from the left, where actions impact the successful execution; for example, not wearing a gas mask before opening the door to a room full of gas would result in player incapacitation and a failed attempt.
The player is notified of the execution outcome through a message box that displays one of three results: no change, failed, or success. Failures do not impact the overall state of the game, allowing the player to modify the plan and attempt again.
Initially, the player is introduced to the game world with a scene containing a conference board with instructions written on it.
The player is instructed on the interface (player, minimap, inventory, planner, and messages), the drag-and-drop mechanic, and how to create scenes and plan actions.
Clicking on the minimap opens a map scene, where the player can choose between three initial locations: the tutorial, the utility closet, and the obstacle.
The player can click on items to get a description of the item in the message system, including hints on how to use it.
New locations are revealed on the map after an obstacle has been successfully cleared, allowing the player to progress.
A final location is unlocked once the player has overcome the gas obstacle twice, once with a gas mask and once with scuba gear.
The last site includes a green button accompanied by a note instructing, "Press the button to complete the game." This was intended to create confusion by departing from the planning system used to overcome earlier challenges.

The play-tests were conducted on premises at the local university.
The experimenter helped the participant with consent, then left them to play solo, checking in and assisting upon request.
Once the player successfully completed the game, they were asked to complete a MiniPXI survey along with their age and gender.
Unfortunately, due to a technical issue that was later fixed, the results of the end survey were lost for a small portion of the participants.

\subsection{Analysis}

Data mining is used to determine the support and confidence of joint affects in the dataset \cite{geng_interestingness_2006}.
Support shows the representativeness of two affects in the dataset.
We set a threshold to 0.1, which means that two affects reported together in at least 10\% of the dataset are considered frequent.
Confidence determines the likelihood that an affect is present in reports featuring a specific affect.
We set the confidence threshold to 0.3, which means that at least 30\% of the reports where the first affect (X) is present, the second affect (Y) is also present.

To test the hypotheses, we use a left-tailed binomial test\cite{dougherty_probability_1990} to support or reject the null hypotheses that players are just as likely to feel (1) frustration and engagement or (2) confusion and boredom as they are to feel either affects in combination with any other affect.
Given that a participant could potentially select between nine different combinations, we set the probability baseline at $1/9$. In the respective case, the alternative hypotheses would then support a probability lower than the probability baseline.

%
To assess the effectiveness of collecting affects during gameplay, we examine their correlation with MiniPXI responses on flow (immersion and mastery) and curiosity in relation to learning experience \cite{dmello_monitoring_2010, dmello_confusion_2014}, which were observed in puzzle games \cite{arguel_puzzle-solving_2019}.
%
This is achievable by defining PXI results as a linear combination of each affect count, utilizing a regression model, typically used for prediction, but also beneficial for data analysis.
%
%
%
Considering each count of affects as random \textit{independent} variables and the MiniPXI constructs as random \textit{dependent} variables, we can use \textit{Bayesian} regression to analyse the influence an independent variable has on a dependent variable.
Bayesian regression quantifies the uncertainty of parameter estimates, with confidence results when it comes to the relation between independent and dependent variables \cite{mcelreath_statistical_2018}.
This approach sees the parameters as the result of probability distributions, rather than fixed points.
Using initial assumptions, known as the \textit{prior distribution}, which is updated with respect to the observable dependent variables (likelihood), we approximate a \textit{posterior} distribution which maximizes the information of the parameters.
%
%
To calculate this, we use Markov Chain Monte Carlo (MCMC)\cite{hastie_elements_2009} by iteratively generating potential posteriors and testing the likelihood of sharing a similar distribution.
After fitting the models, the most appropriate way of interpreting its resulting distribution is through the use of the 94\% High Density Intervals (HDI), which represents the range where the bulk of the posterior probability is concentrated. As a general rule, if 0 is included in the HDI of a coefficient, we say the variable associated with it is non-significant, meaning the affect does not have an influence on the miniPXI response. 
In any other case, we consider the variable significant, making it fair to interpret its distribution and its effect on the dependent variable.

\subsection{Results}

The study included 40 participants aged 20 to 43 (mean 28, SD 6), with 23 males (57.5\%), 6 females (15\%), 3 unspecified (7.5\%), and 8 unknown (20\%). Data on gender, birth year, and miniPXI responses for 8 individuals (20\%) were lost due to technical errors, but other data were not affected. The play time ranged from 7 to 26 min, averaging 13 min 55 s (SD 4 min 13 s). A total of 747 player experience reports were collected, averaging 18.7 per participant (SD 6.6). Single emotional affect was reported 440 times (59\%) and dual affects 307 times (41\%), averaging 11.9 (SD 5.3) for single and 8.3 (SD 6.7) for dual per participant. Reported affects included confusion (400 times, 39 sessions), engagement (301 times, 36 sessions), neutrality (179 times, 34 sessions), frustration (144 times, 28 sessions), and boredom (30 times, 13 sessions).
The neutral affect was reported together with engagement 47 times in 15 sessions with a mean of 3,13 reports (sd 3.09) and confusion 45 times in 16 sessions with a mean of 2.81 reports (sd 1.76).
This is striking in contrast to neutral and frustration, which were reported 7 times in 6 sessions with a mean of 1.17 reports (sd 0.41), or neutral and boredom, which were reported once.
For the sake of simplicity, the neutral affect, when reported together with another affect, is considered similar to not selecting a secondary affect in the rest of the result.
 \begin{table}[]
     \caption{Self report survey statistics.}
     \centering

\begin{tabular}{lrrrr}
Affects & Count & Sessions & Mean & Sd \\
\hline \\
Confusion & 226 & 38 & 5.9474 & 4.4110 \\
Engagement & 175 & 34 & 5.1471 & 4.0161 \\
Confusion \& frustration & 63 & 15 & 4.2000 & 3.8951 \\
Confusion \& engagement & 97 & 25 & 3.8800 & 2.5547 \\
Neutral & 79 & 22 & 3.5909 & 3.1571 \\
Frustration & 49 & 18 & 2.7222 & 3.1212 \\
Boredom & 11 & 5 & 2.2000 & 1.6432 \\
Engagement \& frustration & 28 & 13 & 2.1538 & 1.7246 \\
Boredom \& confusion & 14 & 8 & 1.7500 & 0.8864 \\
Boredom \& frustration & 4 & 3 & 1.3333 & 0.5774 \\
Boredom \& engagement & 1 & 1 & 1.0000 & 0.0000 \\
\end{tabular}

    \label{tab:statistics}
\end{table}

Table \ref{tab:statistics} lists the total self-reported affects, session counts reporting these, and the average and standard deviation per session for single or combined affects in any order.
\begin{table}[]
    \centering
    \caption{Binomial test results for the hypothesis}

\begin{tabular}{llrrrr}
& Label & n & k & Statistic & p \\
\hline \\
H1 & Frustration $\neq$ engagement & 417 & 28 & 0.0671 & 0.0016 \\
H2 & Confusion $\neq$ boredom & 416 & 14 & 0.0337 & \textless0.0001 \\
\end{tabular}

    \label{tab:binomial}
\end{table}
In table \ref{tab:binomial}, $n$ denotes the total number of instances with any affect (including single occurrences), and k indicates where both affects were selected and the p value below 0.05 shows significant evidence to reject the null hypotheses.
This supports the hypothesis that players are unlikely to simultaneously report (H1) both frustration and engagement, or (H2) confusion and boredom.
\begin{table}[]
    \caption{Data mining support and confidence.}
    \centering
    
\begin{tabular}{lrr}
Affects & Support & Confidence \\
\multicolumn{3}{l}{\textbf{Excitatory}} \\
\{Engagement, Confusion\} & 0.1299 & 0.3223 \\
\{Confusion, Engagement\} & 0.1299 & 0.2425 \\
\{Confusion, Frustration\} & 0.0843 & 0.1575 \\
\{Frustration, Confusion\} & 0.0843 & 0.4375 \\
\{Frustration, Boredom\} & 0.0054 & 0.0278 \\
\multicolumn{3}{l}{\textbf{Inhibitory}} \\
\{Engagement, Boredom\} & 0.0013 & 0.0033 \\
\{Engagement, Frustration\} & 0.0375 & 0.0930 \\
\{Frustration, Engagement\} & 0.0375 & 0.1944 \\
\{Confusion, Boredom\} & 0.0187 & 0.0350 \\
\{Boredom, Confusion\} & 0.0187 & 0.4667 \\
\{Boredom, Engagement\} & 0.0013 & 0.0333 \\
\{Boredom, Frustration\} & 0.0054 & 0.1333 \\
\end{tabular}

    \label{tab:datamining}
\end{table}
To analyse whether self-reported affects adhere to the Model of Affect Dynamics, we use data mining to check the support and confidence of pairwise reports, shown in Table \ref{tab:datamining}.
The analysis indicates that engagement and confusion frequently occur with a high level of confidence. 
Furthermore, affect pairs that support the Model of Affect Dynamics appear more frequent with higher confidence than those that oppose it, except in the cases of frustration and boredom.
\begin{figure}
    \centering
    \includegraphics[width=0.8\linewidth]{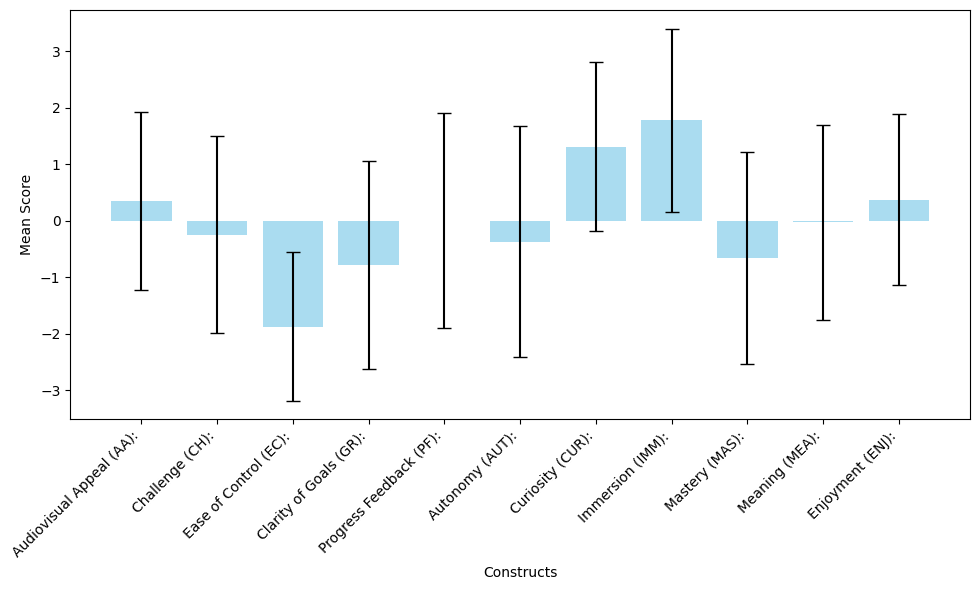}
    \caption{Mean Scores and Standard Deviations of miniPXI Constructs}
    \label{fig:minipxi}
\end{figure}
The results of the MiniPXI (fig. \ref{fig:minipxi}) show high curiosity and immersion, implying that the players were interested and engaged despite the confusing elements of the prototype.
The Bayesian regression models for the PXI constructs of curiosity, immersion, and mastery gave the following results.
We did not find any affect that has a significant influence on the curiosity model, even though boredom came close.
%
In the immersion model, we found that engagement is the only significant affect, with a positive influence, and in the mastery model, engagement has a significant positive influence, while confusion has a significant negative influence.
This supports the relationship with engagement and immersion, as well as the relationship between mastery and engagement from flow theory.

\section{Conclusion and future work}

In this study, we collected subjective experiences of players engaging in a confusing video game and showed that the self-reported experience aligns with the Model of Affect Dynamics.

The results support that 1) video games can be a learning experience, where confusion signals learning and acts as both a predecessor to frustration and a means to prevent boredom, and 2) the experience aligns with the flow experience.
The results are constrained by limited demography, lab setup, and frequent surveys, which may have influenced the reported affects. Thus, more research with more games, demographics, methods of affect collection, and player types is necessary for a more precise understanding of affect dynamics during play.

Future research on player experience regarding flow and learning may be distilled into two trinary scales (1: engaged - neutral - frustrated) and (2: confused - neutral - bored) following the Model of Affect Dynamics. 

\bibliographystyle{plain}
\bibliography{references}

\begin{thebibliography}{10}

\bibitem{abeele_development_2020}
Vero~Vanden Abeele, Katta Spiel, Lennart Nacke, Daniel Johnson, and Kathrin Gerling.
\newblock Development and validation of the player experience inventory: {A} scale to measure player experiences at the level of functional and psychosocial consequences.
\newblock {\em International Journal of Human-Computer Studies}, 135:102370, March 2020.

\bibitem{arguel_puzzle-solving_2019}
Amaël Arguel, Lori Lockyer, Kevin Chai, Mariya Pachman, and Ottmar~V. Lipp.
\newblock Puzzle-{Solving} {Activity} as an {Indicator} of {Epistemic} {Confusion}.
\newblock {\em Frontiers in Psychology}, 10:163, January 2019.

\bibitem{arguel_inside_2017}
Amaël Arguel, Lori Lockyer, Ottmar~V. Lipp, Jason~M. Lodge, and Gregor Kennedy.
\newblock Inside {Out}: {Detecting} {Learners}’ {Confusion} to {Improve} {Interactive} {Digital} {Learning} {Environments}.
\newblock {\em Journal of Educational Computing Research}, 55(4):526--551, July 2017.

\bibitem{baker_towards_2012}
Ryan~S.J.d. Baker, Jessica Kalka, Vincent Aleven, Lisa Rossi, Sujith~M. Gowda, Angela~Z. Wagner, Gail~W. Kusbit, Michael Wixon, Aatish Salvi, and Jaclyn Ocumpaugh.
\newblock Towards {Sensor}-{Free} {Affect} {Detection} in {Cognitive} {Tutor} {Algebra}.
\newblock In {\em Proceedings of the 5th {International} {Conference} on {Educational} {Data} {Mining}}, pages 126--133, 2012.

\bibitem{benmergui_storyteller_2023}
Daniel Benmergui.
\newblock Storyteller, September 2023.

\bibitem{chen_flow_2007}
Jenova Chen.
\newblock Flow in games (and everything else).
\newblock {\em Communications of the ACM}, 50(4):31--34, April 2007.

\bibitem{csikszentmihalyi_toward_2014}
Mihaly Csikszentmihalyi.
\newblock Toward a {Psychology} of {Optimal} {Experience}.
\newblock In {\em Flow and the {Foundations} of {Positive} {Psychology}}, pages 209--226. Springer Netherlands, Dordrecht, 2014.

\bibitem{dougherty_probability_1990}
Edward~R. Dougherty.
\newblock {\em Probability and statistics for the engineering, computing, and physical sciences}.
\newblock Prentice-Hall, Englewood Cliffs, N. J, 1990.

\bibitem{dmello_dynamics_2012}
Sidney D’Mello and Art Graesser.
\newblock Dynamics of affective states during complex learning.
\newblock {\em Learning and Instruction}, 22(2):145--157, April 2012.

\bibitem{dmello_confusion_2014}
Sidney D’Mello, Blair Lehman, Reinhard Pekrun, and Art Graesser.
\newblock Confusion can be beneficial for learning.
\newblock {\em Learning and Instruction}, 29:153--170, February 2014.

\bibitem{dmello_monitoring_2010}
Sidney~K D’Mello and Natalie Person.
\newblock Monitoring {Affect} {States} {During} {Effortful} {Problem} {Solving} {Activities}.
\newblock {\em International Journal of Artificial Intelligence in Education,}, vol. 20(no. 4):361--389, 2010.

\bibitem{geng_interestingness_2006}
Liqiang Geng and Howard~J. Hamilton.
\newblock Interestingness measures for data mining: {A} survey.
\newblock {\em ACM Computing Surveys}, 38(3):9, September 2006.

\bibitem{haider_minipxi_2022}
Aqeel Haider, Casper Harteveld, Daniel Johnson, Max~V. Birk, Regan~L. Mandryk, Magy Seif El-Nasr, Lennart~E. Nacke, Kathrin Gerling, and Vero Vanden~Abeele.
\newblock {miniPXI}: {Development} and {Validation} of an {Eleven}-{Item} {Measure} of the {Player} {Experience} {Inventory}.
\newblock {\em Proceedings of the ACM on Human-Computer Interaction}, 6(CHI PLAY):1--26, October 2022.

\bibitem{hastie_elements_2009}
Trevor Hastie, Robert Tibshirani, and Jerome Friedman.
\newblock {\em The {Elements} of {Statistical} {Learning}}.
\newblock Springer {Series} in {Statistics}. Springer New York, New York, NY, 2009.

\bibitem{hunicke_mda_2004}
Robin Hunicke, Marc LeBlanc, and Robert Zubek.
\newblock {MDA}: {A} {Formal} {Approach} to {Game} {Design} and {Game} {Research}.
\newblock In {\em Proceedings of the {AAAI} {Workshop} on {Challenges} in {Game} {AI}}, Menlo Park, California, 2004. The AAAI Press.

\bibitem{jeanne_ellis_ormrod_educational_2019}
{Jeanne Ellis Ormrod}, {Eric M. Anderman}, and {Lynley H. Anderman}.
\newblock {\em Educational {Psychology}: {Developing} {Learners}}.
\newblock Pearson, 10th edition, 2019.

\bibitem{jonbro_riddle_2006-1}
{JonBro}.
\newblock Riddle {School}, May 2006.

\bibitem{karhulahti_mechanicaesthetic_2011}
Veli-Matti Karhulahti.
\newblock Mechanic/aesthetic videogame genres: adventure and adventure.
\newblock In {\em Proceedings of the 15th {International} {Academic} {MindTrek} {Conference}: {Envisioning} {Future} {Media} {Environments}}, pages 71--74, Tampere Finland, September 2011. ACM.

\bibitem{mcelreath_statistical_2018}
Richard McElreath.
\newblock {\em Statistical {Rethinking}: {A} {Bayesian} {Course} with {Examples} in {R} and {Stan}}.
\newblock Chapman and Hall/CRC, 1 edition, January 2018.

\bibitem{salen_rules_2004}
Katie Salen and Eric Zimmerman.
\newblock {\em Rules of play: game design fundamentals}.
\newblock MIT Press, Cambridge, Mass, 2004.

\bibitem{sweetser_gameflow_2005}
Penelope Sweetser and Peta Wyeth.
\newblock {GameFlow}: a model for evaluating player enjoyment in games.
\newblock {\em Computers in Entertainment}, 3(3):3--3, July 2005.

\bibitem{volden_childrens_2024}
Thomas V.~S. Volden and Paolo Burelli.
\newblock Children’s {Expressed} {Emotions} {During} {Playful} {Learning} {Games}.
\newblock {\em European Conference on Games Based Learning}, 18(1):981--988, October 2024.

\end{thebibliography}

\end{document}